\documentclass[journal,onecolumn]{IEEEtran}

\usepackage{cite}
\usepackage{amsmath}
\usepackage{amssymb}
\usepackage{hyperref}
\usepackage{soul}
\usepackage{amsthm}
\usepackage{xcolor}
\usepackage{caption}
\usepackage{subcaption}
\ifCLASSINFOpdf
\usepackage[pdftex]{graphicx}
\graphicspath{{./pics/}}
\DeclareGraphicsExtensions{.pdf,.jpeg,.png}
\else
\fi

	\usepackage{authblk}
    \title{This is some thing}
    \author[1]{Chang Liu}
    \affil[1]{Department of Information Engineering and Computer Science, University of Trento}
\begin{document}
		
		\title{Exploiting Supply Chain Interdependencies for Stock Return Prediction: A Full-State Graph Convolutional LSTM}

	% make the title area
	\maketitle

\begin{abstract}
Stock return prediction is fundamental to financial decision-making, yet traditional time series models fail to capture the complex interdependencies between companies in modern markets. We propose the \emph{Full-State Graph Convolutional LSTM} (FS-GCLSTM), a novel temporal graph neural network that incorporates value-chain relationships to enhance stock return forecasting. Our approach features two key innovations: First, we represent inter-firm dependencies through value-chain networks, where nodes correspond to companies and edges capture supplier–customer relationships, enabling the model to leverage information beyond historical price data. Second, FS-GCLSTM applies graph convolutions to all LSTM components—current input features, previous hidden states, and cell states—ensuring that spatial information from the value-chain network influences every aspect of the temporal update mechanism. We evaluate FS-GCLSTM on Eurostoxx~600 and S\&P~500 datasets using LSEG value-chain data. While not achieving the lowest traditional prediction errors, FS-GCLSTM consistently delivers superior portfolio performance, attaining the highest annualized returns, Sharpe ratios, and Sortino ratios across both markets. Performance gains are more pronounced in the denser Eurostoxx~600 network, and robustness tests confirm stability across different input sequence lengths, demonstrating the practical value of integrating value-chain data with temporal graph neural networks.
\end{abstract}

	\IEEEpeerreviewmaketitle
	
	\section{Introduction}\label{introduction}

Predicting financial time series has always been a highly sought-after topic among researchers and investors, as it allows for better decision-making in financial markets \cite{GANDHMAL2019100190}. The accuracy of predicted returns is a crucial aspect of any portfolio construction model, as it directly affects the performance and profitability of the portfolio. Historically, research has primarily focused on the techniques of fundamental analysis and technical analysis \cite{FARIASNAZARIO2017115}. However, in recent years, with the increasing availability of computational power, statistical models and machine learning (ML) algorithms have become more prevalent in financial forecasting. These algorithms can analyze large amounts of data and identify patterns that may not be discernible to human traders, allowing for more accurate predictions and better decision-making in financial markets.

For instance, statistical models like autoregressive models (AR), vector autoregression models (VAR), autoregressive integrated moving average models (ARIMA) \cite{7046047}, and heterogeneous autoregressive models (HAR) \cite{har_paper}, in conjunction with multivariate linear regression models \cite{BINI20161248}, are commonly used as benchmarks to be compared with more sophisticated approaches. In fact, with the increasing prevalence of ML and deep learning (DL) models, these sophisticated approaches are becoming more common tools for financial predictions, as researchers seek alternatives to traditional statistical methods that may better capture complex, non-linear patterns in financial data \cite{Zhang_2023}. Researchers have been using these techniques in recent years to replicate the success seen in other areas of research, such as natural language processing and image processing.

However, predicting financial time series poses a unique challenge: compared to ML tasks mentioned above, there is no absolute or objective ``ground truth'' for the stock price. The value of a stock is ultimately determined by the collective beliefs and expectations of all market participants. These beliefs and expectations can be influenced by a wide range of factors, including economic conditions, company performance, news events, and investor sentiment \cite{EFFICIENTCAPITALMARKETS}. Therefore, while predicting stock prices, the ultimate goal is to uncover hidden information from the market to produce excess returns from setting up appropriate investment strategies.

To the best of our knowledge, no prior research has combined temporal features from stock price data with spatial features derived from value chain networks to predict explicit stock returns. In this work, we apply a Full State Graph Convolutional LSTM (FS-GCLSTM) model that integrates LSTM and Graph Convolutional Network (GCN) components. The GCN is used to capture topological relationships from value chain data, modeled as an undirected graph in which each node represents a stock and node features correspond to historical price movements. For each temporal snapshot, GCN layers extract spatial information across the graph. The resulting sequence of spatial representations is then fed into LSTM layers to capture temporal dependencies. Our approach differs from that of \cite{Chen152} in a key architectural detail: we apply GCNs not only to node features but also to all inputs of the LSTM cell, including the previous hidden and cell states. As shown in Section \ref{results}, this design leads to improved prediction performance compared to applying GCNs solely on the node features.

The paper is organized as follows. In Section \ref{model}, we describe the FS-GCLSTM model. In Section \ref{test}, we explain the data and methodology used to test the model. Section \ref{results} presents the empirical results of the study, comparing FS-GCLSTM model to the baseline models and demonstrating its superior properties. We also simulate the model's outcomes to generate a real financial portfolio and compute end-of-period cumulative returns. Finally, Section \ref{outlook} provides concluding remarks.

\section{Previous Research}\label{previous}

Machine learning models have become central to the prediction of financial time series, outperforming traditional statistical methods in many contexts. Classical models such as autoregressive (AR), vector autoregressive (VAR), autoregressive integrated moving average (ARIMA) \cite{7046047}, and heterogeneous autoregressive (HAR) models \cite{har_paper}, often combined with multivariate linear regression \cite{BINI20161248}, remain important benchmarks but are limited in their capacity to capture complex nonlinear patterns in financial data.

Deep learning approaches offer a more flexible framework. Convolutional neural networks (CNNs) are particularly effective in extracting spatial patterns from structured data \cite{726791,10.1145_3065386}. For time-dependent data, recurrent neural networks (RNNs), especially LSTM models, have shown strong performance in modeling sequential dependencies and are widely used for stock return forecasting \cite{8666592,8628641,7364089,Lei,Borovkova}. Hybrid architectures combining CNN and LSTM have also been proposed to benefit from both spatial and temporal feature extraction \cite{Wu,Choi}.

To model interactions between multiple financial entities, graph neural networks (GNNs) provide an effective means of learning from graph-structured data. They enable the incorporation of relationships between stocks or companies into the learning process and are typically classified into spectral and spatial approaches \cite{PAIVA2019635}. Recent works have combined GNNs with temporal models such as LSTM or GRU to jointly capture relational and sequential patterns \cite{DBLP:journals/corr/DefferrardBV16,Seo2016,Ruiz_2020,Chen152,Son}.

In addition to methodological advances, researchers have incorporated non-traditional sources of information to improve predictive accuracy. Sentiment analysis based on news or social media is one such development. Anese et al.~\cite{anese2023impact} investigated the influence of financial news on market returns using LSTM-based classifiers and found that dictionary-based sentiment labels outperformed return-based labels, especially in short time windows.

Alternative financial indicators have also been explored. Grilli and Santoro~\cite{grilli2022forecasting} introduced Boltzmann entropy, derived from agent-based models, as an input to LSTM networks. Their results showed that this entropy measure, used alone or in combination with classical indicators, improved prediction accuracy on both stock and cryptocurrency data while maintaining a relatively simple network architecture.

Network-based representations of financial markets have gained increasing attention. Semenov et al.~\cite{semenov2024weakly} analyzed the structure of graphs formed by weakly correlated stocks, highlighting that even dense graphs often contain small, high-quality cliques. Their comparison of graph construction techniques underlined the importance of accounting for noise in financial correlation networks.

Other studies have modeled company relationships, such as supply chains, industries, or strategic alliances, as graphs to enhance forecasting models. Cohen and Frazzini \cite{Cohen2008} showed that economically linked firms exhibit delayed price responses, suggesting exploitable predictive patterns. Recent works have used such relationships to construct graph structures for learning tasks \cite{WANG2022771,gao_ying,XU2022783,ijcai2020p626,Cheng_Li_2021,8215745,leung_mackinnon}.

This study builds on these findings by integrating temporal price signals with graph-structured value chain information. We propose a model that combines graph convolutional networks with LSTM layers to capture both the structural dependencies between firms and their temporal price evolution. In contrast to previous studies that apply GCNs only to static node features, our architecture propagates graph information into all LSTM inputs, including hidden and cell states. This integration enables the model to learn more expressive representations and improves its ability to anticipate stock returns based on both historical prices and inter-firm relationships.

\section{The FS-GCLSTM Model}\label{model}
We now introduce the architectural details of our proposed model. We propose the \emph{Full-State Graph Convolutional LSTM} (FS-GCLSTM), an extension of the standard GCLSTM in which all LSTM inputs—the previous hidden state $\mathbf{h}_{t-1}$, previous cell state $\mathbf{c}_{t-1}$, and current input features $\mathbf{X}_t$—are first processed by graph convolutional layers before entering the LSTM gates.

\paragraph{Graph Convolutional Layer.}
Let $G=(V,E)$ denote a graph with $n$ nodes, feature matrix $\mathbf{X} \in \mathbb{R}^{n\times d}$, and adjacency matrix $\mathbf{A} \in \mathbb{R}^{n\times n}$, where $\mathbf{A}_{ij}=1$ if $(i,j) \in E$ (or the corresponding edge weight in weighted graphs). The identity matrix $\mathbf{I}_n \in \mathbb{R}^{n\times n}$ is used to add self-loops: $\mathbf{\tilde{A}} = \mathbf{A} + \mathbf{I}_n$, and $\mathbf{\tilde{D}}$ denotes the degree matrix of $\mathbf{\tilde{A}}$. 

Following \cite{kipf2016semi}, the spatial GCN layer is defined as
\begin{equation}
    \mathbf{Z} = \mathbf{\tilde{D}}^{-\frac12} \mathbf{\tilde{A}} \mathbf{\tilde{D}}^{-\frac12} \mathbf{X} \mathbf{W},
\end{equation}
where $\mathbf{W}$ is a learnable weight matrix. A nonlinear activation $f(\cdot)$ (ReLU in our implementation) is then applied:
\begin{equation}
    \mathbf{H} = f(\mathbf{Z}).
\end{equation}
Stacking layers enables multi-hop neighborhood aggregation; in this work, two GCN layers are applied to each processed tensor. We denote this two-layer GCN transformation as $\mathbf{\tilde{H}}(\cdot)$.

\begin{figure}[tb]
    \centering
    \includegraphics[width=5in]{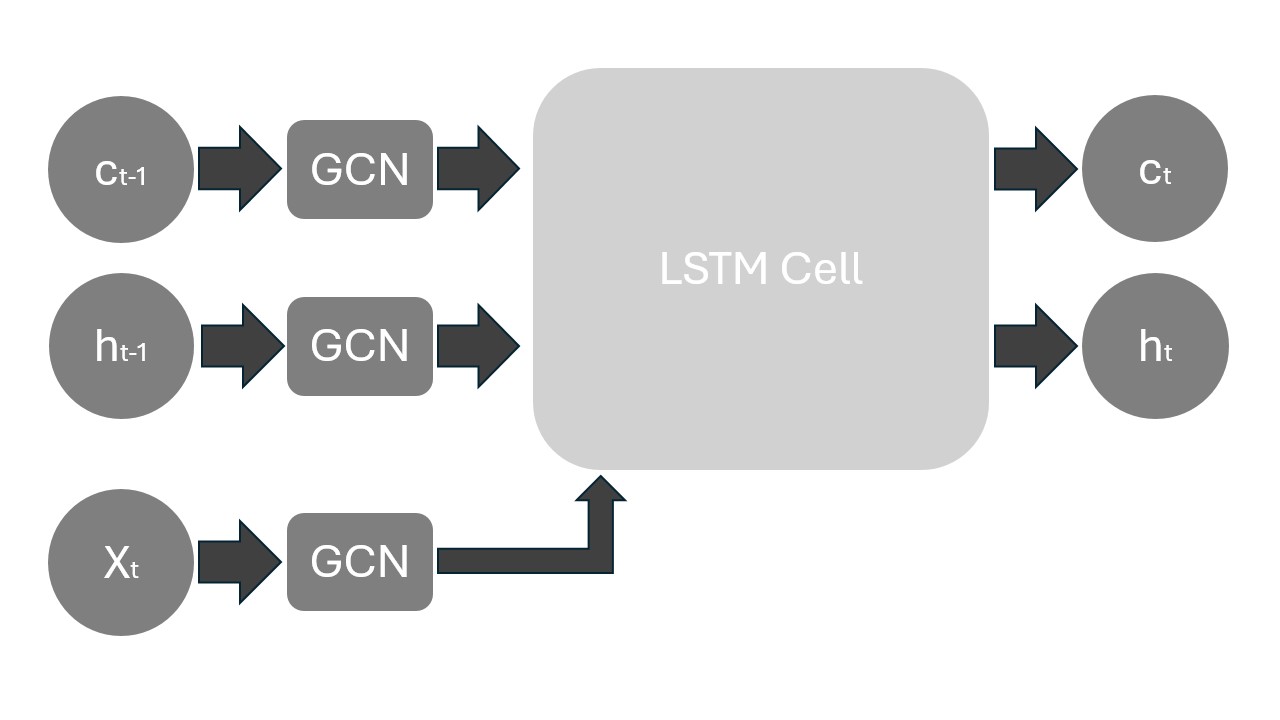}
    \caption{Schematic representation of the FS-GCLSTM cell. The diagram illustrates how GCN layers process the previous hidden state $\mathbf{h}_{t-1}$, previous cell state $\mathbf{c}_{t-1}$, and current input features $\mathbf{X}_t$ before they enter the LSTM gates for the update computations.}
    \label{fig:gclstm_cell}
\end{figure}

\paragraph{FS-GCLSTM Cell.}
At time step $t$, the FS-GCLSTM cell applies $\mathbf{\tilde{H}}(\cdot)$ to $\mathbf{h}_{t-1}$, $\mathbf{c}_{t-1}$, and $\mathbf{X}_t$, as illustrated in Figure~\ref{fig:gclstm_cell}. The transformed tensors are then used in the LSTM update equations:
\begin{align*}
\mathbf{f}_t &= \sigma\!\left(\mathbf{W}_f [\mathbf{\tilde{H}}(\mathbf{h}_{t-1}), \mathbf{\tilde{H}}(\mathbf{X}_t)] + \mathbf{b}_f\right), \\
\mathbf{i}_t &= \sigma\!\left(\mathbf{W}_i [\mathbf{\tilde{H}}(\mathbf{h}_{t-1}), \mathbf{\tilde{H}}(\mathbf{X}_t)] + \mathbf{b}_i\right), \\
\mathbf{c}_t &= \mathbf{f}_t \odot \mathbf{\tilde{H}}(\mathbf{c}_{t-1}) + \mathbf{i}_t \odot \tanh\!\left(\mathbf{W}_c [\mathbf{\tilde{H}}(\mathbf{h}_{t-1}), \mathbf{\tilde{H}}(\mathbf{X}_t)] + \mathbf{b}_c\right), \\
\mathbf{o}_t &= \sigma\!\left(\mathbf{W}_o [\mathbf{\tilde{H}}(\mathbf{h}_{t-1}), \mathbf{\tilde{H}}(\mathbf{X}_t)] + \mathbf{b}_o\right), \\
\mathbf{h}_t &= \mathbf{o}_t \odot \tanh(\mathbf{c}_t),
\end{align*}
where $\sigma(\cdot)$ is the logistic sigmoid activation, $\tanh(\cdot)$ is the hyperbolic tangent activation, $\odot$ denotes the element-wise (Hadamard) product, $\mathbf{W}_*$ are learnable weight matrices, and $\mathbf{b}_*$ are learnable bias vectors. The gates $\mathbf{f}_t$, $\mathbf{i}_t$, and $\mathbf{o}_t$ are the forget, input, and output gates, respectively. As shown in Figure~\ref{fig:gclstm_cell}, this architecture ensures that all inputs to the LSTM gates benefit from the spatial information captured by the graph convolutional layers.

\begin{figure}[tb]
    \centering
    \includegraphics[width=7in]{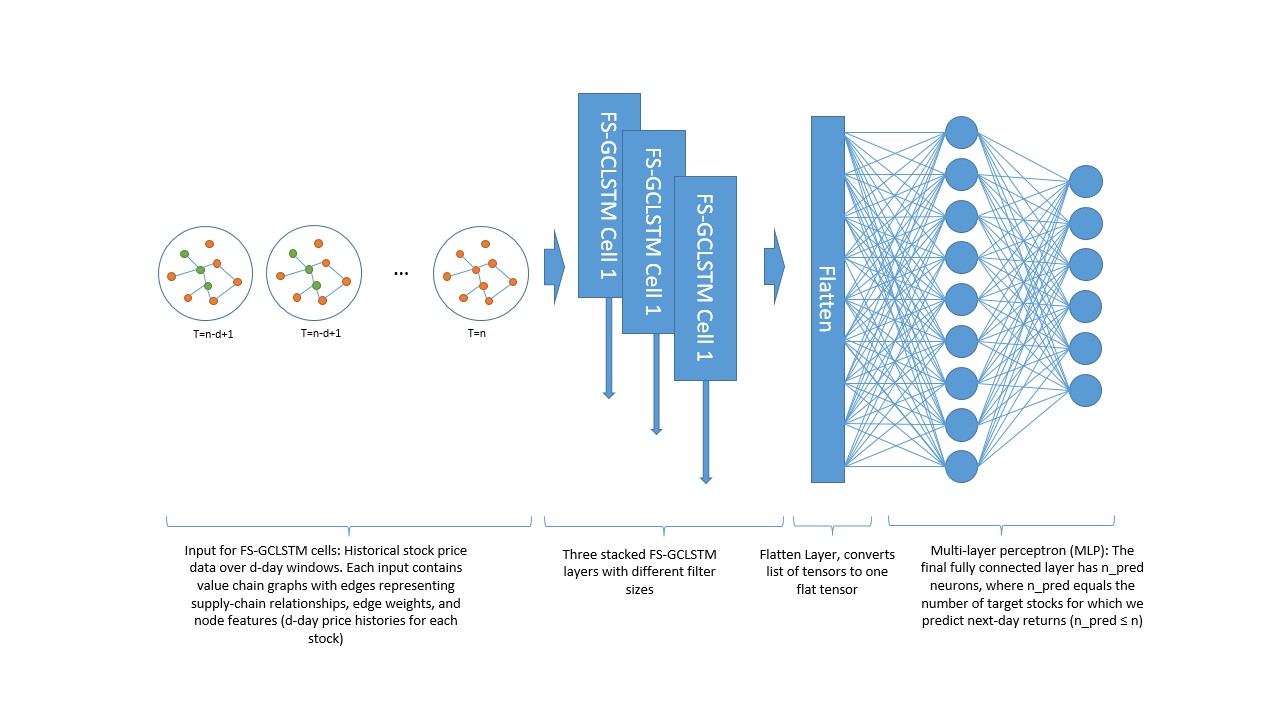}
    \caption{Architecture of the FS-GCLSTM model. Left: Rolling windows of historical stock price data (d trading days) form node features for value chain graphs at each time step. Center: Three stacked FS-GCLSTM cells process the temporal sequence, with each cell applying GCN transformations to all LSTM inputs. Right: Final hidden states are concatenated, flattened, and passed through an MLP to predict next-day returns for selected stocks. The number of output predictions $N_{pred}$ may be different than the total number of nodes in the graph.}
    \label{fig:gclstm_model}
\end{figure}

\paragraph{Model Architecture.}
The overall architecture is illustrated in Figure~\ref{fig:gclstm_model}. We use a rolling window of $d$ trading days as node features, where each node's feature vector $\mathbf{x}_i \in \mathbb{R}^d$ represents the historical price movements for stock $i$. The adjacency matrix $\mathbf{A}_t$ is constructed from value-chain relationships at each time step, where $\mathbf{A}_{ij} = 1$ if companies $i$ and $j$ have a direct supplier-customer connection, and $\mathbf{A}_{ij} = 0$ otherwise. Three stacked FS-GCLSTM cells process the temporal sequence of graphs, with each cell applying GCN transformations to all LSTM inputs as described above. The final hidden states from all three layers are concatenated, flattened, and passed through a multi-layer perceptron (MLP) to predict next-day returns for $N_{\text{pred}}$ selected stocks, where $N_{\text{pred}} \leq n$ may be smaller than the total number of nodes in the graph.

\section{Empirical Analysis}\label{test}
\subsection{Data}

We evaluate the FS-GCLSTM on two major equity markets: the Eurostoxx~600 and the S\&P~500. The Eurostoxx~600 contains 600 large-cap companies from 17 European countries, while the S\&P~500 comprises 500 leading U.S. firms across diverse sectors, including technology, healthcare, and finance.

Inter-firm supplier–customer relationships are obtained from LSEG’s value-chain dataset. For each company, we retrieve its suppliers and customers to construct a directed graph in which nodes represent firms and edges represent value-chain links. Each relationship is assigned a confidence score and timestamp by LSEG, derived from public filings and news sources. Only relationships with a confidence score above $20\%$ are retained in the final graph.

    \begin{table}[!t]
		\caption{Summary of the network properties of Eurostoxx 600 and S\&P500}
		\label{summary_datasets}
		\centering
		% Some packages, such as MDW tools, offer better commands for making tables
		% than the plain LaTeX2e tabular which is used here.
		\begin{tabular}{|c |r| r|}
			\hline
			 & Eurostoxx 600 & S\&P 500\\
			\hline
			\# nodes & 1576 & 1694\\ 
			\# edges & 2501  & 2446\\
            density & 8.105 $10^{-4}$ & 6.513 $10^{-4}$\\
            \# of connected components & 799 & 912\\
            maximum size of connected components & 674 & 638\\
            average size of connected components & 1.9724 & 1.8465\\
			\# features of nodes &  60 & 60\\
            \# output nodes  & 550 & 656\\
			\hline
		\end{tabular}
	\end{table}

\begin{figure}
    \centering
    \begin{subfigure}[b]{0.49\textwidth}
            \includegraphics[width=\linewidth]{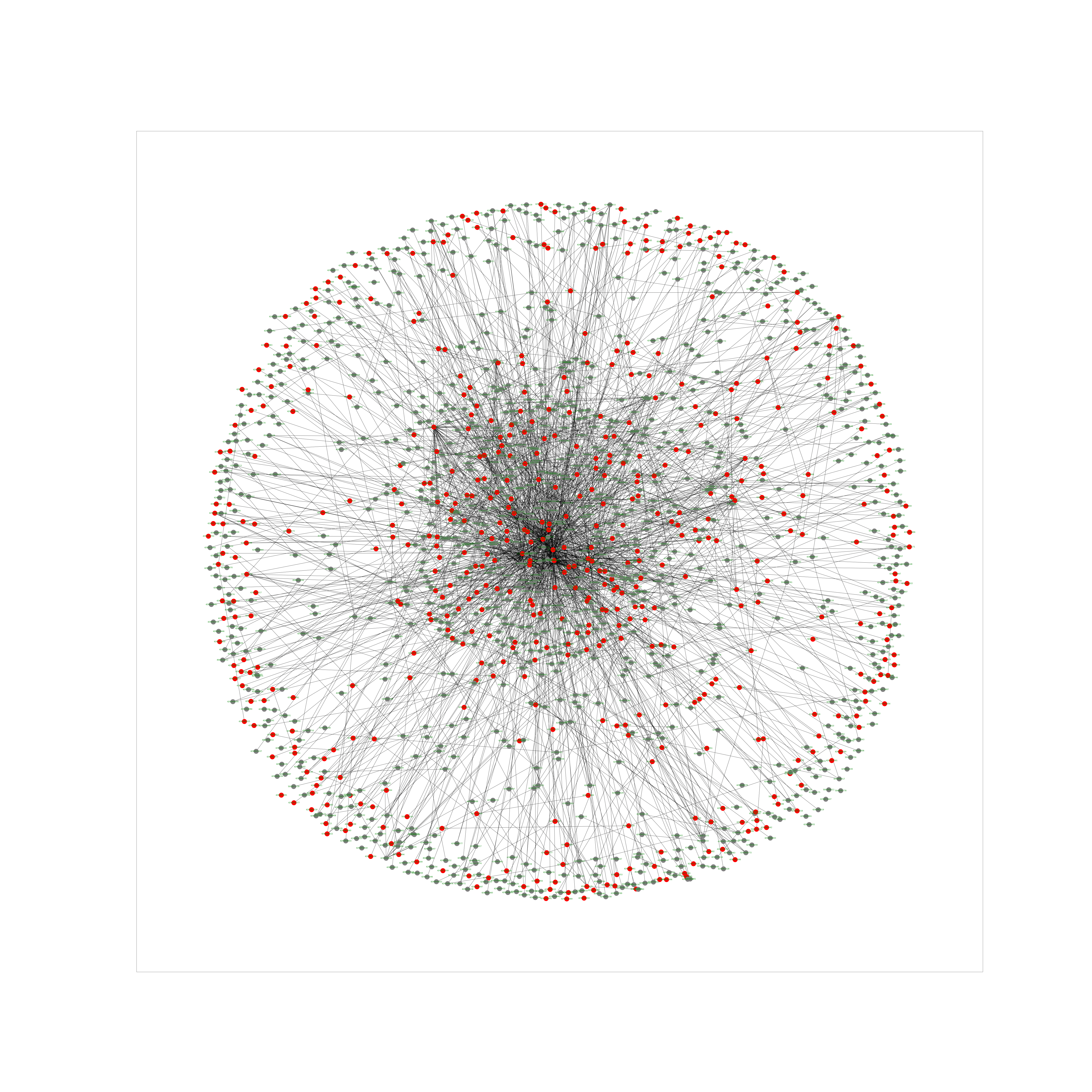}
            \caption{Value chain graph of Eurostoxx 600}
            \label{fig:gull}
    \end{subfigure}
    \begin{subfigure}[b]{0.49\textwidth}
            \includegraphics[width=\linewidth]{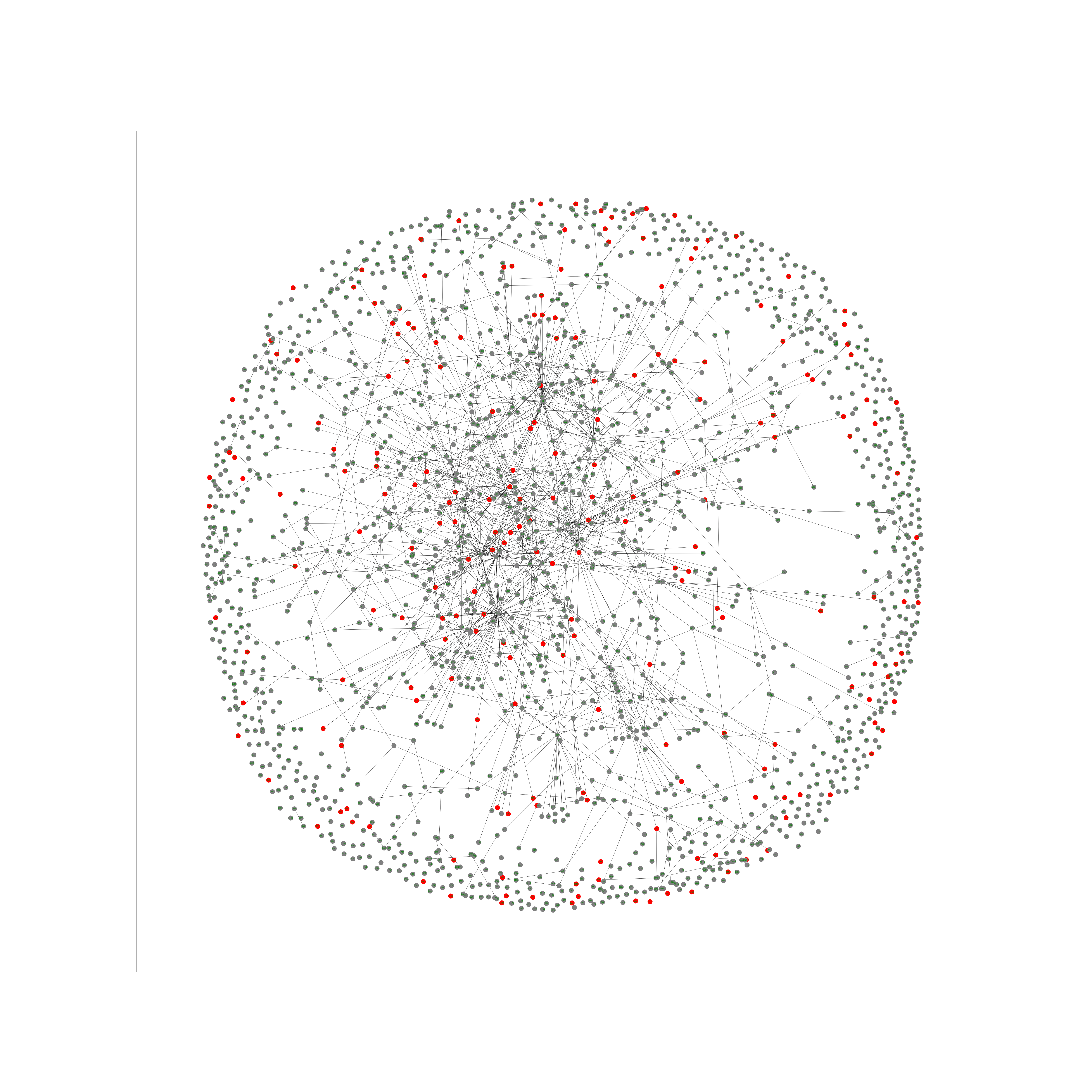}
            \caption{Value chain graph of S\&P 500}
            \label{fig:gull2}
    \end{subfigure}
    \caption{Value chain graphs of Eurostoxx600 and S\&P 500 datasets. Nodes are the listed companies. Red nodes denote companies that are constituents of the corresponding indexes. Edges represent the supplier/customer relationships between two entities. In this layout, components less connected to other components are placed in peripheral space. The graph of Eurostoxx 600 looks denser, although the total number of edges in both graphs is comparable. This is due to the fact that the connections in S\&P value chain data are more concentrated on fewer nodes in the inner space.}
    \label{fig:graphs}
\end{figure}

\subsection{Preprocessing}
To mitigate survivorship bias, we include all companies ever listed in either index and retrieve their historical value-chain data. Firms lacking valid supplier–customer information are removed from the graph. Historical daily closing prices from 2000-01-01 to 2024-12-31 are collected for each firm and its network-linked partners.

We restrict prediction targets to stocks with at least 5{,}000 trading days. For Eurostoxx~600, we retain only firms listed in one of 19 European countries\footnote{Austria, Belgium, Czech Republic, Denmark, Finland, France, Germany, Greece, Ireland, Italy, Luxembourg, the Netherlands, Norway, Poland, Portugal, Spain, Sweden, Switzerland, United Kingdom.}. For the S\&P~500, only U.S.-listed firms are retained. This filtering has a larger impact on Eurostoxx~600 due to the presence of non-European firms in its value-chain network. Consequently, the number of output nodes is smaller than the total number of graph nodes.

As shown in Table~\ref{summary_datasets}, the resulting network structures differ substantially between the two datasets. The Eurostoxx~600 graph exhibits higher density ($8.105 \times 10^{-4}$ vs. $6.513 \times 10^{-4}$), indicating more interconnected value-chain relationships among European firms. Despite having fewer total nodes (1,576 vs. 1,694), the Eurostoxx~600 network contains more edges (2,501 vs. 2,446), reflecting denser supply-chain integration. Both networks are highly fragmented, with the majority of firms organized into small, disconnected components---the Eurostoxx~600 has 799 connected components with an average size of 1.97 nodes, while the S\&P~500 has 912 components averaging 1.85 nodes. However, both datasets feature one dominant connected component containing over 600 firms, representing the core of each region's integrated supply-chain network. The number of prediction targets (output nodes) represents approximately 35\% and 39\% of total nodes for Eurostoxx~600 and S\&P~500, respectively, after applying the geographical and trading history filters.

Figure~\ref{fig:graphs} provides a visual comparison of these structural differences, illustrating the varying density and connectivity patterns between the European and U.S. value-chain networks.

\subsection{Model Set Up} 

We adopt a rolling-window strategy for training and evaluation. The initial window covers 3{,}000 trading days, split into $70\%$ training, $20\%$ validation, and $10\%$ testing. After each iteration, the window advances by 300 days until the dataset is exhausted. Temporal ordering is preserved (no shuffling). Graphs are constructed using the most recent value-chain relationships; edge weights are set to LSEG confidence scores, and edges are treated as bidirectional.

Training uses the Adam optimizer with an initial learning rate of 0.001 and weight decay of $1\times 10^{-5}$. The learning rate is scheduled via OneCycleLR. Models are trained for up to 30~epochs with early stopping after 10~epochs without validation improvement.

\subsection{Baseline Models}

We compare FS-GCLSTM against the following baseline models:
\begin{enumerate}
    \item \textbf{ARIMA:} Rolling window of 60 days; optimal parameters are selected using \texttt{pmdarima} \cite{pmdarima}.
    \item \textbf{FCL:} Four-layer fully connected network with layer sizes $(10\,n_{\mathrm{in}}, 5\,n_{\mathrm{in}}, 10\,n_{\mathrm{in}}, 10\,n_{\mathrm{out}})$, where $n_{\mathrm{in}}$ and $n_{\mathrm{out}}$ denote the numbers of input and output stocks. Input is a flattened $(10, n_{\mathrm{in}})$ return tensor from the last 10 days.
    \item \textbf{LSTM:} Two-layer LSTM with hidden sizes $(60,60,6)$, followed by an MLP with layers $(6\,n_{\mathrm{in}}, 10\,n_{\mathrm{in}}, 1\,n_{\mathrm{out}})$. Cell states are carried forward within the rolling window.
    \item \textbf{GConvGRU:} Chebyshev spectral graph convolutional gated recurrent unit \cite{seo2016structuredsequencemodelinggraph}.
\end{enumerate}

\subsection{Evaluation Metrics}\label{results}
We first assess model performance using mean squared error (MSE), mean absolute error (MAE), and directional correctness (\%), which measure predictive accuracy in a statistical sense. While these metrics are informative, they do not necessarily translate into superior investment performance. For example, overestimating a $+1\%$ return as $+3\%$ may be less harmful to a portfolio than predicting $-1\%$ for the same case, even though the MSE and MAE are identical. To better capture practical utility, we also evaluate the models through portfolio simulations.

The trading strategy is a daily rebalanced, long-only, equal-weighted portfolio: on each day, all stocks with positive predicted returns are equally weighted, with no short selling and transaction costs set to 1~bps. Portfolio allocation is restricted to stocks that were constituents of the corresponding index (Eurostoxx~600 or S\&P~500) at the tradind day to avoid involving penny stocks and ensure practical investability. We compare this to a \emph{constant-weights} benchmark, in which all index constituents are equally weighted at all times. Portfolio performance is evaluated using annualized return, Sharpe ratio \cite{Sharpe49}, and Sortino ratio \cite{Sortino27}. The Sharpe ratio measures excess return per unit of total volatility, while the Sortino ratio considers only downside volatility. Risk-free rates are set to the Euro OverNight Index Average (EONIA) for Eurostoxx~600 and the USD Overnight LIBOR (US00O/N) for S\&P~500.

\begin{figure}[tb]
    \centering 
    \includegraphics[width=7in]{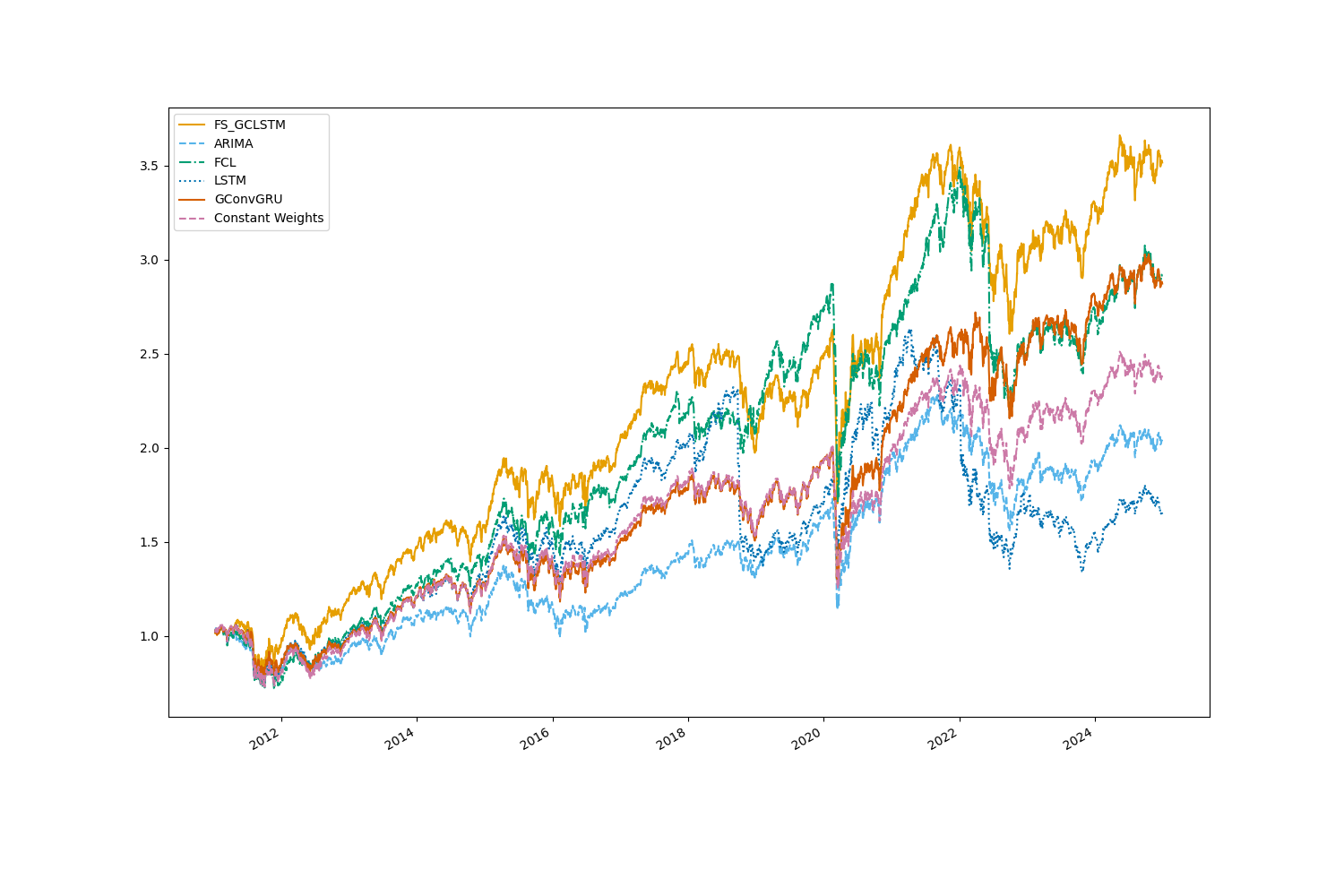} 
    \caption{Cumulative portfolio performance for Eurostoxx~600 using daily rebalanced, equal-weighted, long-only strategies. Portfolio allocation is restricted to current index constituents from the 19 European countries. FS-GCLSTM consistently outperforms all baseline models (ARIMA, FCL, LSTM, GConvGRU) and the constant-weights benchmark throughout the evaluation period.}
    \label{fig:pred1} 
\end{figure}

\begin{figure}[tb]
    \centering 
    \includegraphics[width=7in]{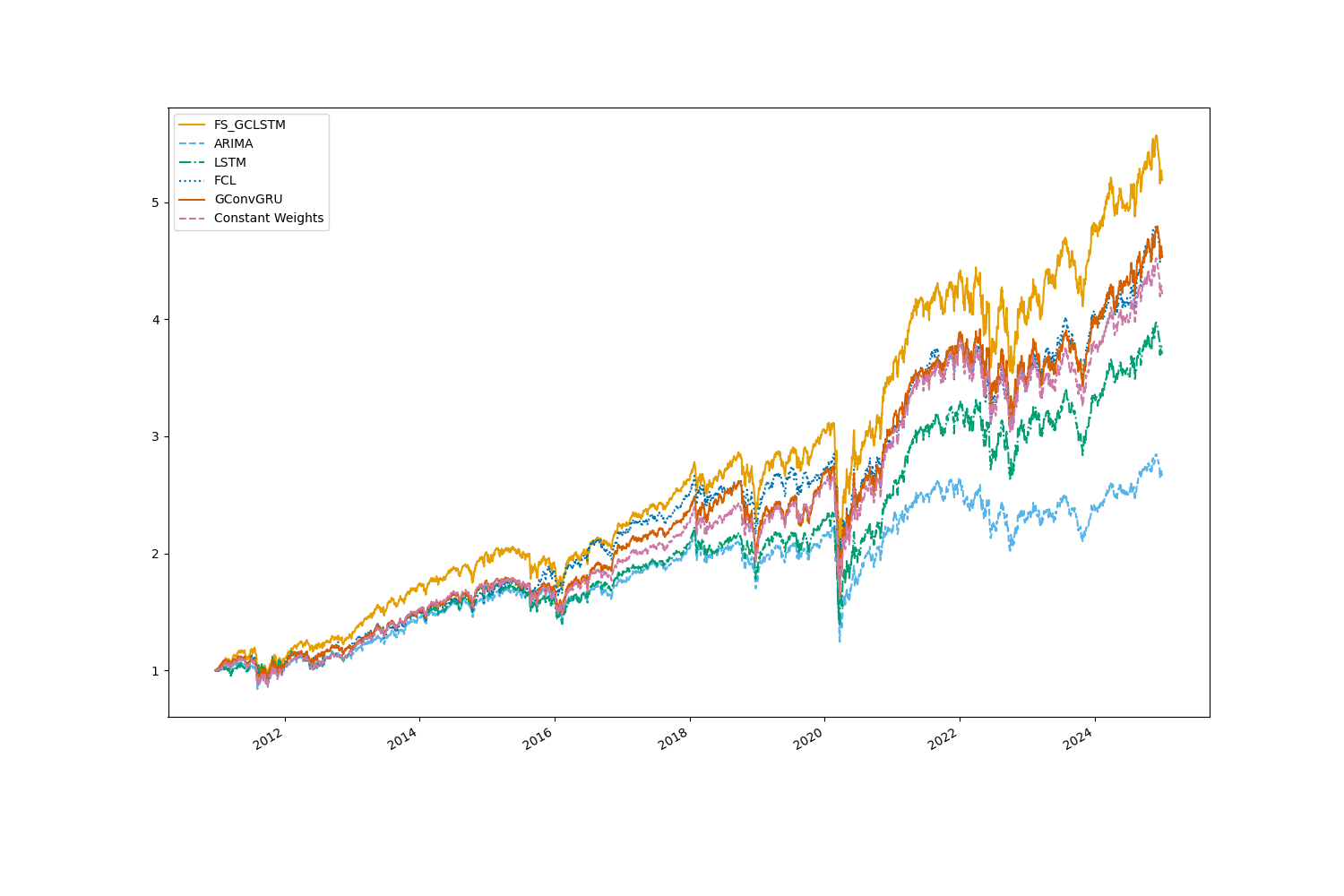} 
    \caption{Cumulative portfolio performance for S\&P~500 using daily rebalanced, equal-weighted, long-only strategies. Portfolio allocation is restricted to current U.S.-listed index constituents. FS-GCLSTM achieves the highest final portfolio value and superior risk-adjusted returns, exceeding all baseline models and the constant-weights benchmark.}
    \label{fig:pred2} 
\end{figure}

Figures~\ref{fig:pred1} and \ref{fig:pred2} present the cumulative performance of all models. FS-GCLSTM consistently delivers the highest portfolio-level metrics across both datasets. In Eurostoxx~600 (Table~\ref{summary_model_comparison_es}), FS-GCLSTM achieves the highest annualized return (7.41\%), Sharpe ratio (0.46), and Sortino ratio (0.59), outperforming all baselines and the constant-weights strategy. In S\&P~500 (Table~\ref{summary_model_comparison_sp}), FS-GCLSTM attains the highest annualized return (9.79\%), Sharpe ratio (0.61), and Sortino ratio (0.75), while its directional correctness is on par with the best baseline models. 

Notably, FS-GCLSTM does not achieve the lowest MSE or MAE in either dataset — those are attained by FCL (Eurostoxx~600) and LSTM (S\&P~500). However, its superior risk-adjusted returns indicate that the model's predictions are more profitable when applied in a trading context, underscoring the importance of portfolio-based evaluation.

\begin{figure}[tb]
    \centering 
    \includegraphics[width=7in]{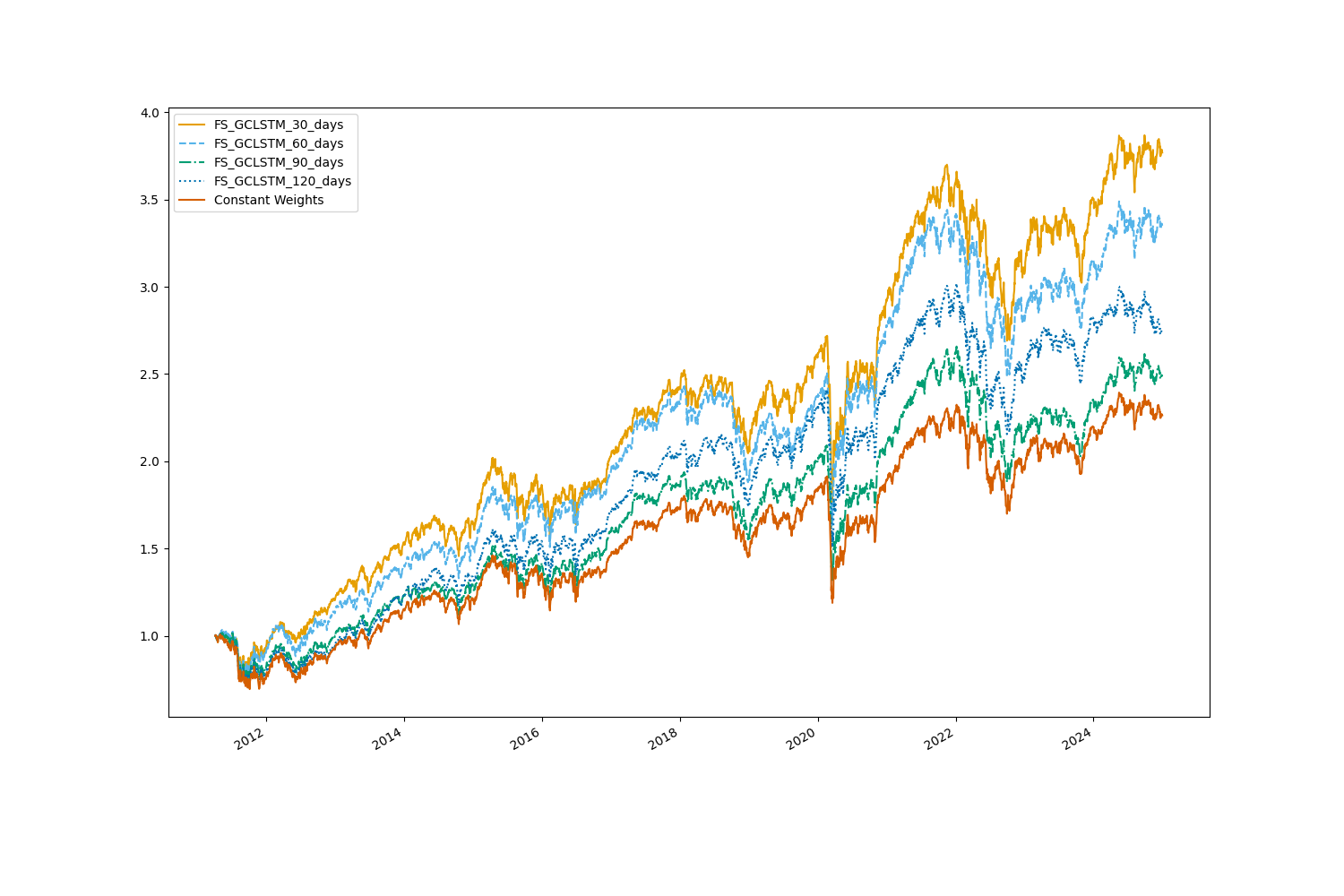} 
    \caption{Robustness analysis for Eurostoxx~600: portfolio performance of FS-GCLSTM across different input sequence lengths (30, 60, 90, and 120 trading days). The model maintains positive risk-adjusted returns across all tested horizons, with optimal performance at 30 days. Performance metrics include annualized return, Sharpe ratio, and Sortino ratio.}
    \label{fig:robust_es600} 
\end{figure}

\begin{figure}[tb]
    \centering 
    \includegraphics[width=7in]{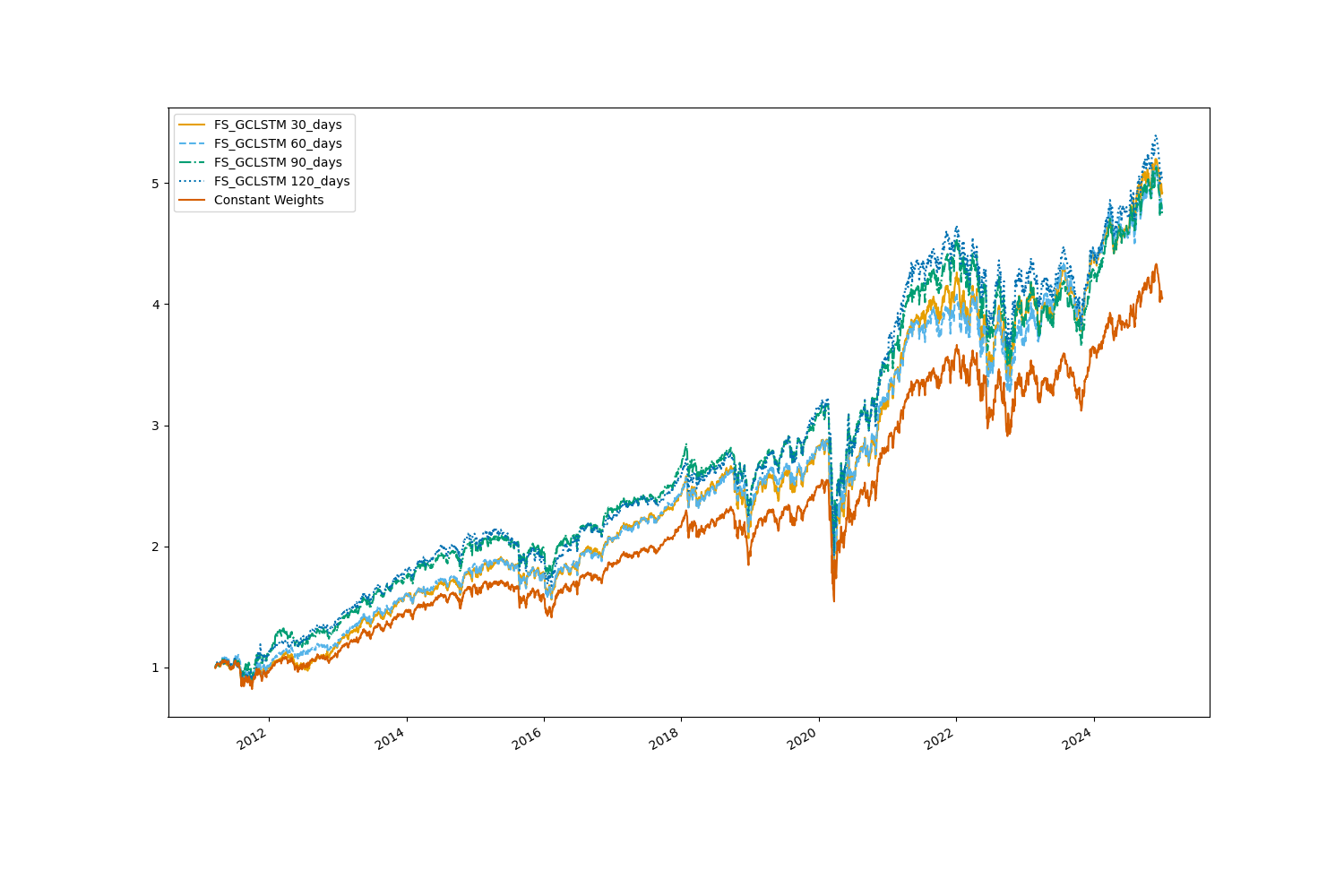} 
    \caption{Robustness analysis for S\&P~500: portfolio performance of FS-GCLSTM across different input sequence lengths (30, 60, 90, and 120 trading days). The model demonstrates stable and consistently high performance across all tested horizons, maintaining superior risk-adjusted returns compared to the constant-weights benchmark.}
    \label{fig:robust_sp500} 
\end{figure}

We further assess robustness by varying the input sequence length from the default 60 days to 30, 90, and 120 days. For Eurostoxx~600 (Table~\ref{tbl:robust_es600}, Fig.~\ref{fig:robust_es600}), FS-GCLSTM maintains positive annualized returns, Sharpe, and Sortino ratios for all tested lengths, with performance peaking at 30 days and declining moderately for longer horizons. For S\&P~500 (Table~\ref{tbl:robust_sp500}, Fig.~\ref{fig:robust_sp500}), performance remains stable across all tested lengths, consistently outperforming the constant-weights benchmark.

	% \begin{table}[!t]
	% 	\caption{Comparison of Models for Eurostoxx 600}
	% 	\label{summary_model_comparison_es}
	% 	\centering
	% 	\begin{tabular}{|c |c| c| r|  r|  r|  r|}
	% 		\hline
 %            models & MSE &  MAE &  Correctness (\%) &ann. Return (\%) & ann. Sharpe Ratio & ann. Sortino Ratio\\
 %            \hline
 %            ARIMA           & 4.408 $10^{-4}$        & 1.388053 $10^{-2}$      & 49.671491         & 4.1211         &  0.2603           & 0.3293\\
 %            FCL             & 4.0814 $10^{-4}$        & 1.34735 $10^{-2}$      & 52.897788         & 6.2422         &   0.3857           & 0.4666\\
 %            LSTM            & 4.1108 $10^{-4}$        & 1.357539 $10^{-2}$      & 52.76455         & 2.8772         & 0.1505           & 0.1636\\
 %            GConvGRU          & 4.3218 $10^{-4}$        & 1.4081939 $10^{-2}$       & 50.665542         &  6.1793      &   0.3938           &   0.51\\
 %            FS-GCLSTM        & 4.2357 $10^{-4}$        & 1.379287 $10^{-2}$     & 50.589807        & \textbf{7.4062}         &   \textbf{ 0.4615}           &  \textbf{0.592}\\
 %            Constant Weights & - & - & - & 5.0345 & 0.3095 & 0.3918\\
	% 		\hline
	% 	\end{tabular}
	% \end{table}

    \begin{table}[!t]
	\caption{Comparison of Models for Eurostoxx 600}
	\label{summary_model_comparison_es}
	\centering
	\begin{tabular}{|c |c| c| r|  r|  r|  r|}
		\hline
		models & MSE &  MAE &  Correctness (\%) &ann. Return (\%) & ann. Sharpe Ratio & ann. Sortino Ratio\\
		\hline
		ARIMA           & 4.408 $\times 10^{-4}$        & 1.388 $\times 10^{-2}$      & 49.67         & 4.12         &  0.260           & 0.329\\
		FCL             & \textbf{4.081} $\times 10^{-4}$        & \textbf{1.347} $\times 10^{-2}$      & \textbf{52.90}         & 6.24         &   0.386           & 0.467\\
		LSTM            & 4.111 $\times 10^{-4}$        & 1.358 $\times 10^{-2}$      & 52.76         & 2.88         & 0.151           & 0.164\\
		GConvGRU        & 4.322 $\times 10^{-4}$        & 1.408 $\times 10^{-2}$       & 50.67         &  6.18      &   0.394           &   0.510\\
		FS-GCLSTM       & 4.236 $\times 10^{-4}$        & 1.379 $\times 10^{-2}$     & 50.59        & \textbf{7.41}         &   \textbf{0.462}           &  \textbf{0.592}\\
		Constant Weights & - & - & - & 5.03 & 0.310 & 0.392\\
		\hline
	\end{tabular}
\end{table}
	
	% \begin{table}[!t]
	% 	\caption{Comparison of Models for S\&P 500}
	% 	\label{summary_model_comparison_sp}
	% 	\centering
	% 	\begin{tabular}{|c |c| c| r|  r|  r|  r|}
	% 		\hline
 %            models & MSE &  MAE &  Correctness (\%) &ann. Return (\%) & ann. Sharpe Ratio & ann. Sortino Ratio\\
 %            \hline
 %            ARIMA           & 4.1003 $10^{-4}$        & 1.308 $10^{-2}$      & 48.7349        & 5.7043         &  0.3588           & 0.4306\\
 %            FCL             & 4.2639 $10^{-4}$        & 1.389 $10^{-2}$      & 50.3937         &   7.7138         &   0.458           &   0.5677\\
 %            LSTM            & 3.8574 $10^{-4}$        & 1.2851 $10^{-2}$      & 50.0266         & 8.9301         & 0.5494           & 0.6829\\
 %            GConvGRU          & 3.9731 $10^{-4}$        & 1.3198 $10^{-2}$       & 49.9295         &  8.9527      &   0.542           & 0.6715\\
 %            FS-GCLSTM        & 3.9846 $10^{-4}$        & 1.3146 $10^{-2}$     & 50.3545        & \textbf{9.7877}         &   \textbf{0.6082}           &  \textbf{0.7535}\\
 %            Constant Weights & - & - & - & 8.516 & 0.5292 & 0.6561\\
	% 		\hline
	% 	\end{tabular}
	% \end{table}	

    \begin{table}[!t]
	\caption{Comparison of Models for S\&P 500}
	\label{summary_model_comparison_sp}
	\centering
	\begin{tabular}{|c |c| c| r|  r|  r|  r|}
		\hline
		models & MSE &  MAE &  Correctness (\%) &ann. Return (\%) & ann. Sharpe Ratio & ann. Sortino Ratio\\
		\hline
		ARIMA           & 4.100 $\times 10^{-4}$        & 1.308 $\times 10^{-2}$      & 48.73        & 5.70         &  0.359           & 0.431\\
		FCL             & 4.264 $\times 10^{-4}$        & 1.389 $\times 10^{-2}$      & 50.39         &   7.71         &   0.458           &   0.568\\
		LSTM            & \textbf{3.857} $\times 10^{-4}$        & \textbf{1.285} $\times 10^{-2}$      & 50.03         & 8.93         & 0.549           & 0.683\\
		GConvGRU        & 3.973 $\times 10^{-4}$        & 1.320 $\times 10^{-2}$       & 49.93         &  8.95      &   0.542           & 0.672\\
		FS-GCLSTM       & 3.985 $\times 10^{-4}$        & 1.315 $\times 10^{-2}$     & \textbf{50.35}        & \textbf{9.79}         &   \textbf{0.608}           &  \textbf{0.754}\\
		Constant Weights & - & - & - & 8.52 & 0.529 & 0.656\\
		\hline
	\end{tabular}
\end{table}	

 %    \begin{table}[!t]
	% 	\caption{Robustness Test for Eurostoxx 600}
	% 	\label{tbl: robust_es600}
	% 	\centering
	% 	\begin{tabular}{|c |c| c| r|  r|  r|  r|}
	% 		\hline
 %            input days length  & MSE &  MAE &  Correctness (\%) &ann. Return (\%) & ann. Sharpe Ratio & ann. Sortino Ratio\\
 %            \hline
 %            30 & 42352 $10^{-4}$        & 1374945 $10^{-2}$      & 51.198241        &    7.9708           &  0.5142         & 0.6583             \\
 %            60 &  4.2357 $10^{-4}$        & 1.379287 $10^{-2}$     & 50.589807        &    7.2392  & 0.4488 & 0.5743     \\
 %            90 & 4.2824 $10^{-4}$        & 1.381845 $10^{-2}$      & 50.679472        &  5.4045            & 0.3361          & 0.427             \\
 %            120 & 4.3316 $10^{-4}$        & 1.398229 $10^{-2}$      & 50.308891        & 6.0085            &0.3682          & 0.4662             \\
	% 		\hline
	% 	\end{tabular}
	% \end{table}	

\begin{table}[!t]
	\caption{Robustness Test for Eurostoxx 600}
	\label{tbl:robust_es600}
	\centering
	\begin{tabular}{|c |c| c| r|  r|  r|  r|}
		\hline
		input days length  & MSE &  MAE &  Correctness (\%) &ann. Return (\%) & ann. Sharpe Ratio & ann. Sortino Ratio\\
		\hline
		30 & 4.235 $\times 10^{-4}$        & 1.375 $\times 10^{-2}$      & \textbf{51.20}        &    \textbf{7.97}           &  \textbf{0.514}         & \textbf{0.658}             \\
		60 &  \textbf{4.236} $\times 10^{-4}$        & 1.379 $\times 10^{-2}$     & 50.59        &    7.24  & 0.449 & 0.574     \\
		90 & 4.282 $\times 10^{-4}$        & 1.382 $\times 10^{-2}$      & 50.68        &  5.40            & 0.336          & 0.427             \\
		120 & 4.332 $\times 10^{-4}$        & \textbf{1.398} $\times 10^{-2}$      & 50.31        & 6.01            &0.368          & 0.466             \\
		\hline
	\end{tabular}
\end{table}

 %    \begin{table}[!t]
	% 	\caption{Robustness Test for S\&P 500}
	% 	\label{tbl: robust_sp500}
	% 	\centering
	% 	\begin{tabular}{|c |c| c| r|  r|  r|  r|}
	% 		\hline
 %            input days length & MSE &  MAE &  Correctness (\%) &ann. Return (\%) & ann. Sharpe Ratio & ann. Sortino Ratio\\
 %            \hline
 %            30 & 3.9685 $10^{-4}$        & 1.310844 $10^{-2}$      & 50.44056        &  9.6218           &  0.5862         & 0.7286             \\
 %            60 &  3.9846 $10^{-4}$        & 1.3146 $10^{-2}$     & 50.3545        &  9.4574  & 0.5857 & 0.7243     \\
 %            90 &  4.0203 $10^{-4}$        & 1.324944 $10^{-2}$      & 50.379874        & 9.4204            & 0.5839          & 0.7308             \\
 %            120 & 4.094 $10^{-4}$        & 1.339136 $10^{-2}$      & 50.064207        & 9.7543            & 0.5901          & 0.7371             \\
	% 		\hline
	% 	\end{tabular}
	% \end{table}	

\begin{table}[!t]
	\caption{Robustness Test for S\&P 500}
	\label{tbl:robust_sp500}
	\centering
	\begin{tabular}{|c |c| c| r|  r|  r|  r|}
		\hline
		input days length & MSE &  MAE &  Correctness (\%) &ann. Return (\%) & ann. Sharpe Ratio & ann. Sortino Ratio\\
		\hline
		30 & \textbf{3.969} $\times 10^{-4}$        & \textbf{1.311} $\times 10^{-2}$      & \textbf{50.44}        &  9.62           &  0.586         & 0.729             \\
		60 &  3.985 $\times 10^{-4}$        & 1.315 $\times 10^{-2}$     & 50.35        &  9.46  & 0.586 & 0.724     \\
		90 &  4.020 $\times 10^{-4}$        & 1.325 $\times 10^{-2}$      & 50.38        & 9.42            & 0.584          & 0.731             \\
		120 & 4.094 $\times 10^{-4}$        & 1.339 $\times 10^{-2}$      & 50.06        & \textbf{9.75}            & \textbf{0.590}          & \textbf{0.737}             \\
		\hline
	\end{tabular}
\end{table}
	
\section{Conclusion}\label{outlook}

In this paper, we introduced the \emph{Full-State Graph Convolutional LSTM} (FS-GCLSTM), a novel neural network architecture for stock return prediction that integrates graph convolutional networks (GCNs) and long short-term memory (LSTM) networks. The model represents value-chain relationships as graphs, where nodes correspond to companies, edges represent supplier–customer links, and node features are derived from historical stock performance. The key innovation of FS-GCLSTM lies in applying graph convolutions not only to the current node features but also to the previous hidden and cell states at each time step, ensuring that spatial information from the value-chain network influences all components of the LSTM update mechanism.

We evaluated FS-GCLSTM on two major financial markets using Eurostoxx~600 and S\&P~500 datasets, comparing it against several baseline models including ARIMA, fully connected layers, LSTM, and graph convolutional GRU. Our empirical results demonstrate that while FS-GCLSTM did not achieve the lowest prediction errors in traditional statistical metrics (MSE and MAE), it consistently outperformed all baselines in economically meaningful portfolio-level performance measures. In both markets, FS-GCLSTM delivered the highest annualized returns, Sharpe ratios, and Sortino ratios when applied in a daily rebalanced, equal-weighted, long-only trading strategy restricted to index constituents.

The model's superior performance was more pronounced in the Eurostoxx~600 market, where the denser and more uniformly connected value-chain network structure enhanced the model's ability to extract informative spatial features. In the sparser S\&P~500 network, FS-GCLSTM still achieved consistent gains over benchmarks, though with smaller margins, suggesting that network density and connectivity patterns significantly influence the effectiveness of value-chain information in financial prediction tasks.

Robustness analysis across different input sequence lengths (30, 60, 90, and 120 trading days) confirmed the stability of our approach. For Eurostoxx~600, performance peaked at shorter horizons (30 days) and declined moderately for longer sequences, while S\&P~500 performance remained consistently high across all tested lengths, indicating robust generalization capabilities.

Our findings demonstrate that temporal graph neural networks applied to value-chain data can successfully uncover profitable trading opportunities, with performance gains varying systematically across different market structures. The practical significance of our results is underscored by the consistent outperformance in risk-adjusted returns, which translates directly to superior investment outcomes.

Several promising avenues exist for extending this work. First, incorporating additional graph modalities such as price-similarity networks or sentiment-based connections could provide complementary information to value-chain relationships. Second, constructing heterogeneous graphs that include entities beyond listed companies—such as private suppliers, investment managers, or regulatory bodies—could capture broader economic dependencies. Finally, moving beyond single-tier supply chain networks to multi-tier value chains would enable the model to capture indirect relationships and cascading effects that propagate through multiple levels of supplier-customer connections.
	\bibliographystyle{IEEEtran}
	
	\bibliography{ref.bib}
	
\end{document}